# Can a short intervention focused on gravitational waves and quantum physics improve students' understanding and attitude?

**Rahul K. Choudhary**[1], **Alexander Foppoli**[1]**, Tejinder Kaur**[1]**, David G. Blair**[1]**, Marjan Zadnik**[1]**, Richard Meagher**[2]

[1]*The University of Western Australia, 35 Stirling Highway, Crawley, WA 6009, Australia.*
[2]*Mount Lawley Senior High School, Mount Lawley, WA 6050, Australia*

Email: rkc.girija@gmail.com

**Abstract**
The decline in students' interest in science and technology is a major concern in the western world. One approach to reversing this decline is to introduce modern physics concepts much earlier in the school curriculum. We have used the context of the recent discoveries of gravitational waves to test benefits of one-day interventions, in which students are introduced to the ongoing nature of scientific discovery, as well as the fundamental concepts of quantum physics and gravitation, which underpin these discoveries. Our innovative approach combines role-playing, model demonstrations, single photon interference and gravitational wave detection, plus simple experiments designed to emphasize the quantum interpretation of interference. We compare understanding and attitudes through pre and post testing on four age groups (school years 7, 8, 9 and 10), and compare results with those of longer interventions with Year 9. Results indicate that neither prior knowledge nor age are significant factors in student understanding of the core concepts of Einsteinian physics. However we find that the short interventions are insufficient to enable students to comprehend more derived concepts.

## 1. Introduction

The recent discovery of gravitational waves has been described as "the discovery of the century".[1] The momentous discoveries from 2015 to 2017 proved the existence of gravitational waves as ripples in space-time predicted by Einstein [2] and opened the new field of gravitational astronomy.[3] They also proved that gravity travels at the speed of light,[4] and provided definitive observations of black holes.[3] Finally they proved that heavy elements are created in the coalescence of neutron stars.[5]

Gravitational waves can only be understood in the context of general relativity while the extremely sensitive detectors which detected these waves can only be understood in the context of quantum mechanics. In particular, the sensitivity of these detectors is achieved by a careful balance of quantum effects associated with the random arrival of photons, and the random forces imparted by the photons when they reflect off the mirrors used in the detectors.

To comprehend gravitational waves, students need to understand the fundamental concepts of Einsteinian physics: space-time as an elastic medium, and quantum mechanics, the key to which is understanding the particle nature of light. Thus the discovery of gravitational waves provide a perfect context for learning quantum ideas of Einsteinian physics and interpret the modern understanding of light.

There is an increasing recognition of the need to modernize school physics. This has led to the development of projects like ReleQuant, [6] the curriculum of Excellence in Scotland [7] and curriculum in Korea, [8] which introduces these topics to high school students and LIGO-EPO [9] which has strongly emphasized the importance of gravitational waves to the general public. Research has shown that it is both possible and effective to introduce Einsteinian physics early in schools.[10][11] The research reported here is a part of the "Einstein-First" project carried out by a team of physicists and science education researchers in Western Australia. We are developing and testing materials based on models and analogies [11] [12] [13] designed for introducing Einsteinian physics at an early age.

A particular reason for early introduction of Einsteinian physics concepts is to avoid conceptual conflicts that arise when Newtonian and Euclidean concepts that have been taught implicitly or explicitly at school must be replaced later by Einsteinian concepts. For example Euclidean geometry and associated geometrical formulae taught at primary school carry with them implicit belief of mathematical exactness while in reality they are approximations. Newtonian gravity carries with it the idea that space is an abstract mathematical entity independent of matter, time is absolute and gravity is a force field that emerges from planets. In reality, spacetime is more like an elastic fabric, time is relative, and gravity is a manifestation of curved spacetime, and in particular the gradient in time created by mass. If the Einsteinian paradigm can be learnt from an early age, students will be able to experience a seamless progression of learning from primary to tertiary level. We do not advocate abandoning the classical physics. However it should be understood as a set of useful and important approximations that can be introduced after learning the Einsteinian paradigm.

Another reason for introducing Einsteinian physics at an early age is that much of Einsteinian physics, from quantum mechanics of semiconductors and cameras, to the gravitational time dilation in GPS clocks, is embodied in mobile phones which gives a level of relevance that is important for student motivation.

The Einstein-First project has developed a set of models and analogies which when tested in longer interventions has yielded positive results. In a 20-lesson program, results show that students can achieve substantial understanding and long term retention of core concepts. [14] Such long interventions are difficult to implement within school curriculum. For this reason, we investigate the understanding and attitudinal changes of students during a single day intervention. This allowed us to compare four different age groups with identical programs.

Short interventions have been shown to be significant in influencing student's career choices. [15] For example, in a survey carried out on over 1000 students by Wynarczyk and Hale, it was found that 52% of the participants said their career choices were influenced by a scientists' or engineers' visit to their workplace, and 24% of them commented that they chose their career by the influence of their visit to scientists' and engineers' workplaces. [16] This assertion is resonated by a quote based on short interventions from S. Laursen et al.: [15]

"When the intervention strategy chosen is short in duration, like a scientist's visit to a classroom, the immediate outcomes of such events are primarily affective, compared with other strategies such as curriculum change or teacher professional development where deeper learning may take place. Short duration intervention strategies are based on a change model with the premise that developing interest and enthusiasm around science, having positive experiences with science, meeting science role models, and learning about science careers will translate down the road to more students pursuing advanced science education and careers in high school, college, and beyond."

As mentioned earlier, there is a decline in the number of students studying science.[17] A study by G. Hasan on 1745 secondary and university students across Australia, suggested declining intrinsic motivation among secondary students and a need to investigate innovative approaches to teaching science.[17]

The above concerns and conclusions motivated us to design a short intervention program that combines multiple components by including a visit to a gravitational wave research centre; interactions with researchers; role plays to explore the ongoing process of discovery; whole class activities to explore the physics of gravitational wave detectors and small group experiments in which students record laser interference measurements using their mobile phones.

To investigate the age response of students, we delivered identical programs to four age groups from the same school (years 7, 8, 9 and 10) of academically talented students. We used identical pre and post tests to determine their initial knowledge, and attitudes. This age range was designed to span the grade level and age when students' interest in science begin to decline.[18] We discover what level of understanding the students have in advance, particularly to assess their prior knowledge and ability to comprehend concepts which are only taught at senior specialist level. The findings are analyzed and compared with longer interventions conducted previously in this project with Year 9 students.

In the following section, we summarize the physics context of our program: gravitational wave detectors and the quantum description of light including a discussion, and the core and derived quantum concepts we wish to impart. In Section 3, we describe the activity based learning methods we used including the role play, the models and analogies, and the hands-on experiment sessions. In Section 4, we describe the research methodology. Section 5 presents research findings and results which include interesting gender effects and the differing responses to core and derived concepts. Results are compared with the results of longer but similar interventions. This allows us to draw clear conclusions in regard to the benefits of short interventions discussed in the final section.

## 2. Gravitational wave detectors and the quantum properties of light

In this program, we used video materials to first illustrate the idea that gravitational waves, ripples in space-time, change the distances between pairs of objects because the space itself stretches and shrinks oppositely in perpendicular directions. We use this concept, and the exciting discovery gravitational waves from coalescing black holes and neutron stars as a motivation, but focus on the concept of ultra-precision measurement using an interferometer. We contrast the classical description of light shown in a LIGO video [19] with the quantum description based on photons, which naturally explains how measurement noise arises due to the discrete nature of photons. This is illustrated using videos of single photon interference. [20] The contrasting pictures of interference illustrated in Figure 1 become the main focus of our program.

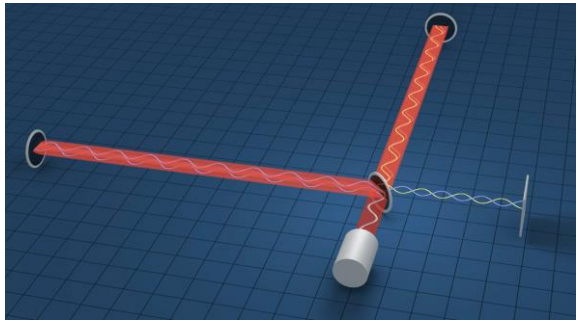 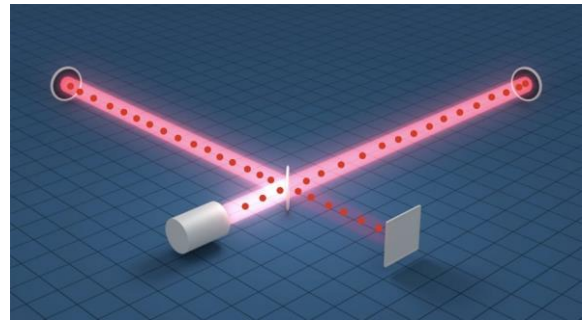

Figure 1(a): A screen grab from the LIGO video that uses a classical picture of light to explain how a Michelson interferometer is sensitive to the stretching and squeezing of space. A classical wave splits exactly in half at a beam splitter, and later recombines.
Figure 1(b) When classical light is replaced by photons (represented by dots) the nature of quantum interference must be confronted, but measurement noise and measurement uncertainty due to the statistical nature of light follows naturally.

Our program is based on Richard Feynman's qualitative vector approach introduced in *QED: The Strange Theory of Light and Matter* (1985). [21] We introduce Feynman through the role play, with his assertion:

*"I want to emphasize that light comes in this form – particles. It is very important to know that light behaves like particles, especially for those of you who have gone to school, where you were probably told something about light behaving like waves".*

Modern optical technology makes it easy to observe the quantum nature of light, but quantum interference was recognized in 1909, when G. I. Taylor showed that a 2000 hour exposure of highly attenuated X-rays still produces an interference pattern.[22] It has recently been shown that human eyes can sense single photons.[23] Videos of single particle interference show that the apparent wave nature only emerges when large numbers of photons create a quasi-continuous pattern characteristic of a classical wave. [24] [13]

Our program replaces the classical interpretation of diffraction and interference with the idea that whenever light is allowed to take two alternative paths, then the phenomenon of interference occurs. It is a phenomenon in which the probability of a photon arriving at a particular location follows the mathematics of vectors, and depends on a property shared by all moving objects: wavelength.

In our program, we visualise the concept of wavelength in ocean waves (we show Google Earth images of ocean waves diffracting around islands) but emphasise that it is a property of all matter and radiation. Similarly momentum (which we experience when we catch a fast ball) is also a universal property of all matter and radiation. Everything has wavelength and everything has momentum.

For advanced students (but not used in this one day intervention) the universal relationship between wavelength and momentum can be introduced. This is the De-Broglie wavelength formula

$$\text{wavelength} = \text{Planck's constant}/\text{momentum}.$$

The tiny magnitude of Planck's constant tells us that wavelength is important for tiny things but unimportant for cricket balls. Many modern physics experiments use interference of electrons, neutrons, atoms and molecules. All are characterized by a wavelength which depends on the particle's momentum.

In quantum mechanics, the wave nature is normally described by wave functions which are mathematical entities used to determine the photon arrival probabilities. The wave function is a mathematical tool that allows the mathematics of adding up classical waves to be used to calculate the probability of particles arriving at a particular location. At the high school level, this mathematical formulation can be represented by introducing simple vector arithmetic.

To introduce the concepts of quantum mechanics in a one day intervention is challenging. In order to encapsulate the quantum concepts, we have broken them down into four concepts, each supported by observational facts.

(a) **Momentum**: *Light comes as photons which like all moving objects, carry momentum. They can transfer momentum to the object they strike*. In particular the momentum of photons pushes the mirrors in gravitational wave detectors, and thereby creates small disturbances.

(b) **Wavelength**: *Photons have a property called wavelength like you can see in ripples on a pond. When photons can take two alternative paths, the probability of photons arriving may be high or low depending on the difference in path length between the two paths. This is called interference.* (see Figure 2) This concept is emphasized by making a link between videos of single photon interference and interference images that we see with the naked eye.

(c) **Randomness**: *Quantum processes are intrinsically random.* Photons arriving at a beam splitter randomly choose one direction or the other, but with a precisely defined probability. Single photon interference demonstrates the intrinsic randomness of photons arriving, but image eventually builds up to form a precise pattern of interference. The small wavelength of visible light (less than a micron) determines the scale of interference and this property can be used to measure very small distances.

(d) **Vector Mathematics**: *The calculation of the probability of photons arriving is described by a new sort of mathematics called vectors, the addition of arrows*. Whereas with numbers $1 + 1 = 2$, if arrows are opposed you can have $1 + 1 = 0$. In interference, the bright fringes are the regions where the arrows are aligned. Dark fringes are the regions where the arrows are opposed. In our one day program we consider only the simplest cases such as

$$\uparrow + \uparrow = \uparrow$$
$$\uparrow + \downarrow = 0$$

In our program the connection between vector mathematics, wavelength and interference, and between the classical and quantum descriptions, is made by using transparent cards with sine wave patterns. Students can align cards to create waves in phase and in anti-phase. This allows intuitive recognition of the role of wavelength: that a half wavelength displacement is sufficient to take you from $1+1 = 2$ to $1+1 = 0$. Hence the arrow directions are associated with the phase difference between the waves.

But in the quantum regime, when only one photon may be present at a time, it is clear that the interference is a property of the path and the wavelength, and is not due to two particles somehow cancelling each other. This is weird but it is the nature of reality. This is exactly the same mathematics that you have to use when classical waves interfere. The smaller the wavelength the more sensitive you are to differences in length of trajectory.

For advanced students we can generalize from photons to electrons, neutrons etc. and introduce imaging with electron microscopes which make images at the atomic scale for which the wavelength of light is too large. Qualitative vector mathematics can be extended to multiple alternative path, and students can grasp the connection between the vector direction and the phase of the wave.

Feynman extends this qualitative vector approach to note that alternative paths can also be arbitrary differences in trajectory, such as curved paths. He uses vector summation of multiple alternative paths to explain why light travels in straight lines and why for mirrors the angle of incidence equals the angle of reflection. These concepts were beyond the capacity of a short intervention but can be explored in longer interventions.

Our program was to determine what level of quantum understanding we could impart in a one day intervention. Our overriding goal was to introduce two core concepts: light comes as photons, and photons carry momentum. In addition we wanted to measure the uptake the three derived concepts: (a) interference occurs whenever the light can take two alternative paths (b) momentum of photons causes measurement uncertainties and (c) wavelength sets the scale size for interference.

### 3. Active learning methods for quantum properties of light

Active (i.e. hands-on) learning has been shown to be effective in promoting critical thinking. [25] Therefore, we designed a program which engages students to learn quantum properties of light through a succession of activities and experiments in the context of gravitational waves. As discussed in section 1, it includes role play, activities based on models and analogies, videos and images, face to face discussions, and a hands-on experiment session.

*3.1 Science role play*
The program begins with a role play session designed to create a learning environment that creates laughter and encourages whole class involvement.

The role play brings together three key scientists Heinrich Hertz, Albert Einstein and Richard Feynman as well as a journalist and a narrator. Students wear simple costume props to identify their roles. No preparation is required: students read a simple script with highlighted text for each actor (See Appendix 1).

Simple props indicate Hertz's laboratory environment. The role play explains the discovery of electromagnetic waves by Hertz and the emergence of quantum description of electromagnetic radiation. It depicts how in 1905 Einstein used the concept of photons to explain Hertz' observation of the photoelectric effect in 1887, [26] [27] and how both the photon concept and Einstein's prediction of gravitational waves were endorsed and confirmed by Feynman. The role play is designed to establish a parallel between electromagnetic waves and gravitational waves, leading to expectations of future breakthroughs as the gravitational wave spectrum is explored. The entire theme of the role-play encapsulates the ongoing process of scientific enquiry which leads to successive questions, predictions, discoveries and further questions.

*3.2 Learning with a Toy Model Interferometer*

To prepare students for interference experiments, we use a toy model interferometer to allow interactive whole class experiments and discussion, in which several volunteers undertake simple tasks that illustrate the key concepts discussed in section 2.

3.2.1 Radiation pressure: We use toy-model photons and suspended toy mirrors (see Figure 3) to demonstrate the effect of photon momentum on mirrors used in gravitational wave interferometers. The toy photons are nerf-gun bullets so the nerf-gun is a toy laser. Students observe the recoil of the mirrors due to the momentum of the nerf-gun bullets. Lighter mass mirrors recoil more than more massive mirrors. The recoil can be estimated using a ruler to determine the relationship between recoil and mirror mass. The main concept of this activity demonstrates that the random arrival of photons causes uncertainty in the position of the mirrors. This is a manifestation of the Heisenberg uncertainly principle in quantum mechanics. The effect of photon momentum acting on mirrors in gravitational wave detectors sets limits on the sensitivity of gravitational wave detectors.

3.2.2 Creating alternative paths for photons: To help students learn the key concept that interferometers require light to take two alternative paths to reach the same point, we give students the practical challenge of creating two alternative paths for the toy interferometer. They are presented with a flat plastic sheet (such as a CD case) which acts a partially reflecting mirror for the light from a green laser pointer. Partial reflection means that photons have a probability of being reflected or transmitted. Students working in teams or in front of the class with other students advising, are challenged to direct beams simultaneously towards the two mirrors of the toy interferometer shown in Figure 3. Most students experience significant difficulties but quickly learn the key aspects of reflected and transmitted beams that enables them to complete the challenge by correctly orienting the beam splitter.

3.2.3 Single photon interference and simple vector addition: Supported by videos and images of laser interferometer gravitational wave detectors such as Figure 1, combined with videos of single photon interference, [28]we emphasize that interference patterns arise because of changes in the probability of the arrival of photons.

In dark locations the probability is zero, while in bright locations the probability is 1. Vector mathematics is introduced using wooden arrows. Students are asked to add them to create a value of 2 or 0. Students quickly grasp the idea of placing arrows head to tail to create different resultants which include arbitrary resultants between 0 and 2 if they are oriented in different angles.

The important connection between vector direction and physical waves is made through use of transparent cards displaying the sine function as discussed in section 2. The key messages are (a) the mathematics of the addition of classical waves is the same as the mathematics of adding quantum probabilities, and (b) one wavelength equals one cycle of rotation of vector direction. [29] The connection with the macroscopic world is emphasized using Google Earth images of ocean waves diffracting around an island (See Figure 4(a)), which itself prepares students for the experiment of the diffraction of laser light around a human hair described in 3.5(c) below.

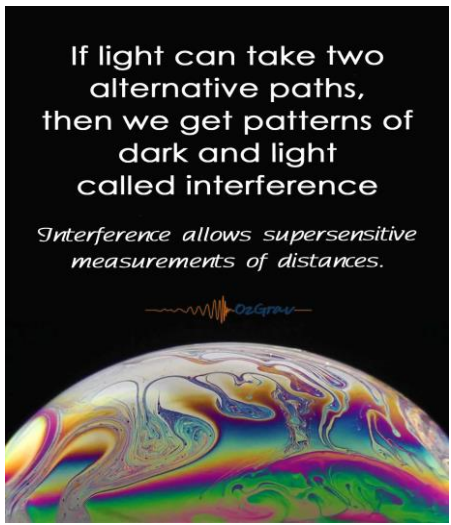 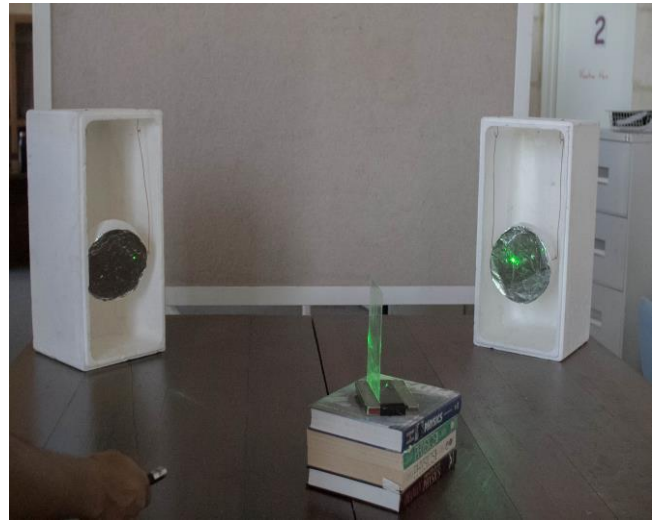

Figure 2. A poster used to depict the concept of interference arising whenever light can take two (or more) alternative paths. The soap bubble colour patterns are an everyday example of interference.

Figure 3. The toy model interferometer activity. Nerf gun bullets are fired at the mirrors to illustrate the concept of photon momentum. Then a beam of green laser light is split by a transparent plastic plate into two beams directed towards the two polystyrene suspended mirrors placed perpendicular to each other. The students are challenged to achieve the laser beam alignment which requires careful adjustment of the beam splitter and understanding of reflection and transmission.

*3.4 Real research experience and laboratory tour*

As mentioned earlier, students' career choices are influenced by direct interaction with researchers and scientists. The purpose of this activity is to allow students to meet young researchers (Ph.D students and researchers working at the Australian International Gravitational-wave Observatory at Gingin, Western Australia), to realize that researchers are ordinary people like themselves, and to hear about the fun and the difficulties of scientific research.

*3.5 Three experiments on photon interference*

The program described below is based on three sets of three low-cost laser interferometer experiments designed to allow 18 students to work together in pairs. The experiments use low cost laser modules. Only one of them, the Michelson interferometer, uses optical equipment in the form of surface coated mirrors, a beam splitter, mirror mounts and a rigid frame. The experiments are designed to reinforce the conceptual framework introduced earlier. We make a particular effort to remind students that because light travels extremely fast, there is generally only one photon present in the apparatus at any time, so the experiments show phenomena similar to the single photon interference they observed in a video. Students use mobile phones to capture images and sound, and do a simple calculation for the experiments.

(a) *Model gravitational wave detector with visible and acoustic output*: A tabletop Michelson interferometer was designed to demonstrate how gravitational wave detectors use interference to make super-sensitive and precise measurements (See Figure 6(a) and 6(b)). The toy interferometer activity described in section 3.2 establishes the idea of splitting a beam, but now students can observe re-combination of the beam by adjusting alignment screws. The interference pattern is observed on a screen, and a photodiode readout (by sampling the output beam using a piece of window glass to pick-off some of the output light) allows audio frequency fringe pattern movement to be heard on a loudspeaker. By gently tapping the table we hear fluctuations in the interference pattern similar to those created by gravitational waves passing through gravitational wave interferometers. Ears can hear these fast changes which cannot be seen on the screen because our eyes respond much more slowly. However slow changes such as those caused by pressing on the table are easily seen. This instrument is very sensitive and students can hear the changes caused by the slightest touch on the table.

The small-scale interferometer illustrates the physics of gravitational wave detectors, but the latter are about 1 billion times more sensitive. The increase in sensitivity is achieved by using very high intensity light. The high intensity is needed because the random arrival of photons (as seen in single photon interference videos) is another source of uncertainty. Based on the same statistical principles as opinion poll testing, the measurement precision improves inversely as the square root of the number of photons. At the laser light power levels used (hundreds of kW corresponding to $10^{25}$ photons per second) the radiation pressure effects (see section 3.21) balance the statistical uncertainties in photon arrivals. This allows motions as small as $10^{-19}$ m to be detected.[30] Sensitivity is limited by the quantum statistics of the photons.

(b) Quantum weirdness in soap film interference: Soap film interference represents an everyday case where under normal conditions there is generally less than one photon present at a time in the region of interference. Because a soap film is so thin, $\sim 10^{-6}$ m, we need about $10^{14}$ photons per second (speed/thickness) to have an average of one photon present at any instant (See Figure 5) for illustration of the interference process, as seen classically and when light is a stream of photons. It is difficult to imagine these photons, therefore we use pictures to illustrate the principle of interference.

When a laser beam is allowed to pass through a soap film, reflections from the upper and rear surface produces interference patterns as shown in Figure 7(a) and 7(b). The interference takes place because of the phase difference caused by the upper and rear surface of the soap film. This experiment is another vivid example of quantum interference taking place due to light travelling into two different paths. Because the soap film is so thin, only one photon at a time is present in the soap film. [13]

(c) Measuring the thickness of human hair: This simple experiment can be performed with any simple laser but is most easily seen with a green laser. This experiment is supported by images of ocean waves diffracting past islands, (See Figure 4(a)) to make the connection between macroscopic ocean wave diffraction and optical diffraction. Light diffracted by the hair produces an interference pattern on a screen (see Figure 4(b) for set up). The two alternative paths on either side of the hair have path differences dependent on the position on the screen. The fringe spacing depends on the thickness of the hair so it allows students to measure their own hair thickness and compare with other class members. [See Appendix 2]

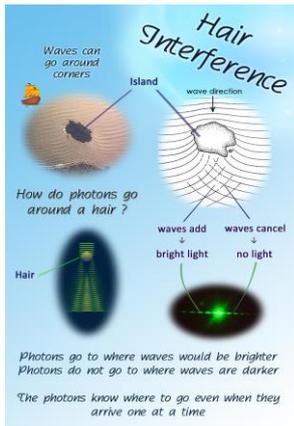 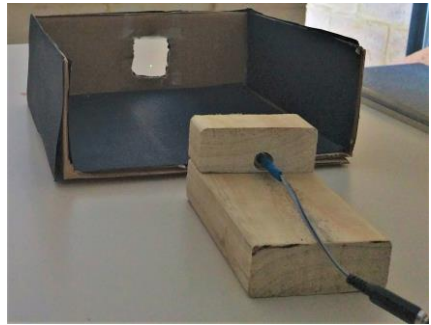 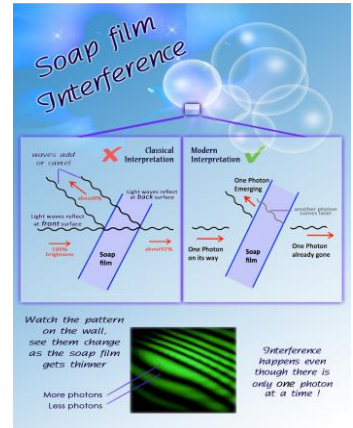

Figure 4 (a) A poster of human hair diffraction. The Google earth image showing waves diffracting around an island. This image is shown and compared with the diffraction of light by human. It is used as an analogy to show how diffraction is a characteristics of both matter and radiation.

Figure 4(b) A laser beam strikes a vertical human hair (held by double sided tape) across a window in a cardboard frame. The diffraction pattern created by a vertical hair. The distance between consecutive bright or dark fringes is used to calculate the hair diameter using the formula given in Appendix 2

Figure 5. A poster used in the program to interpret the soap film interference. We contrast the classical and quantum descriptions of light.

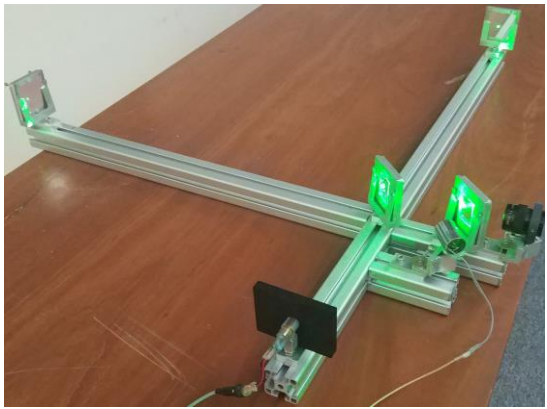 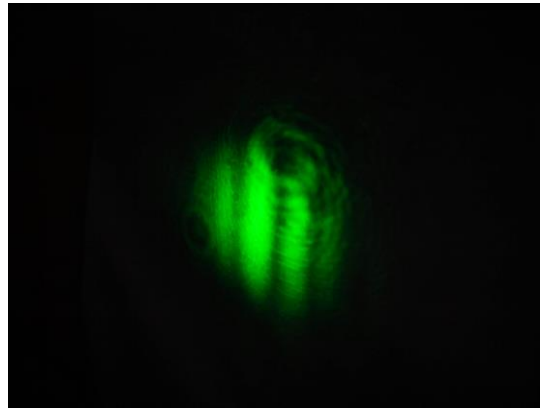

Figure 6(a) A table-top Michelson interferometer used to investigate the physics of gravitational wave detectors. The beam splitter, end mirrors and pick off glass for audio output are illuminated.

Figure 6(b) Interference pattern observed on a wall about 1 meter from the Michelson interferometer output.

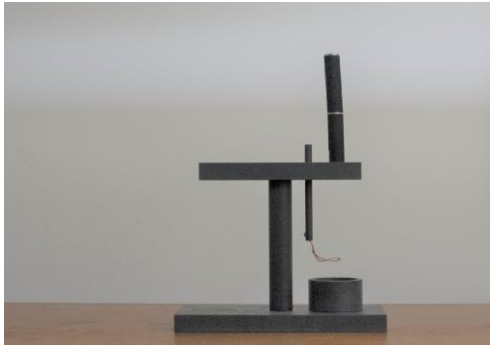 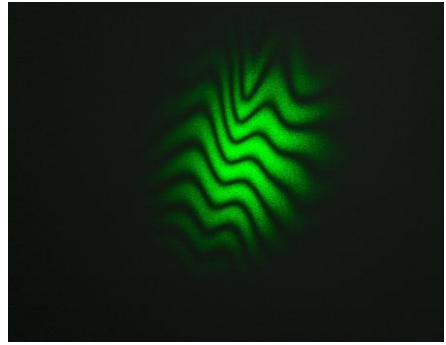

Figure 7(a) Set up for soap film interference. A laser directed downwards reflects from an angled soap film is created by dipping a copper wire loop into a pot of soap solution. Out-of-plane twisting of the loop is creates a spherical film surfaces that magnifies the reflected beam.
Figure (b) The high contrast interference pattern projected onto a nearby wall. Dark regions are places where the probability of photons arriving due to interference is minimal and the bright regions are the places where the probability is maximum. The pattern evolves with time and also can display many beautiful effects such

## 4. Methodology

*4.1 Participants and program delivery:* The intervention program reported here involved 103 students from Year 7 to 10 of Mount Lawley Senior High School, Western Australia. The students of Year 7 were gifted and talented in language while the students of Year 8-10 were gifted and talented in science. The programs were delivered by the same group of researchers, and students from the same school were selected deliberately to minimize any effects of socio-economic background.

Our program does not use control groups to compare results. Rather we compare our results with previous longer interventions, and between age groups. We select middle school age groups because we want to test ability to accept Einsteinian concepts before the contradictory Newtonian concepts have become too entrenched. We wanted to determine the level of acceptance as a function of age, thereby helping us determine how to structure an optimum curriculum that would build an Einsteinian concept understanding throughout student's schooling.

*4.2 Questionnaire description:* We developed two pairs of identical questionnaires: pre/post program conceptual understanding and pre/post program attitudinal tests. The pre-program tests were conducted in the classroom before arrival at the Australian International Gravitational Observatory site, while post-program tests were also conducted in the classroom after a gap of one day. Respondents were allotted 15-20 minutes for their responses.

*4.2.1 Conceptual questionnaire:* The conceptual questions were developed mainly considering two basic elements: (a) ability to understand the core concepts and (b) ability to understand the derived concepts discussed in Section 2. For example "Can light exert forces on things?" was used as a question to test students' understanding that photons carry momentum which is one of the core ideas of Einsteinian physics. The question "What aspect of light can cause uncertainty in measurement?" termed as "derived concept" was used to test students' ability to understand the phenomenon of uncertainty principle by applying this concept into a new situation.

In total seven questions (see Table 2) were set in advance for the conceptual understanding test. (Q1) & (Q3) tested core concepts and (Q4), (Q5) & (Q7) tested derived concepts. (Q2) asked students to name three different types of electromagnetic waves and explain what they are, designed to test students' ability to remember and explain scientific facts from the program. (Q6) tested the concept that light has energy, a familiar concept in a country where solar energy is very popular. (Q7) was related to a derived Einsteinian concept that energy has mass. (Q6) & (Q7) also served as an insight to students' prior knowledge.

These conceptual questions required short-responses, the answers to which were evaluated and marked. The marking criteria were full marks for the correct answer with explanation, half for correct answer without explanation and 0 for no response or incorrect response. Every question had different weightage viz. Q1, Q3, Q4, Q5 carried 2 marks, Q6 &Q7 carried 1 marks and Q2 carried 3 marks.

*4.2.2 Attitudinal questionnaire:* There were seven attitudinal questions based on Likert scale items. The Likert scale marks were quantified according to 1– 5 scale, 1 for strongly disagree and 5 for strongly agree and inverted for negative questions. We calculate the average score for each question. (See Table 3)

*4.2.3 Questionnaire validity and interpretation:* The questions were validated on the basis of following important points:

(a) We used the topics covered in the program as the basis in designing conceptual and attitudinal questionnaires, where only content-related items were tested.
(b) It was considered whether the students are able to interpret the questions correctly. For example, for the question "What is light", it was explicitly explained that here "light" does not mean the opposite of "heavy".
(c) The attitudinal questions were selected from the literature, [31] [32] drafted and modified to match the program.
(d) The questionnaires were verified by a panel of science educators and researchers for their validity.

*4.3 Data analysis*
This research was framed within a quantitative, empirical-analytical design. Data analysis of conceptual questionnaire was done by comparing student's pre and post program scores. Students' responses to each question was analyzed in terms of gender and age. The ratio of post and pre-test average scores were obtained to find the improvement factor for each attitudinal question. Scores of Question no. 3 and 7 in attitudinal questionnaires were inverted for analysis because they were negative in nature. Furthermore, t-tests were performed to determine statistically significant difference in scores with p-value set as 0.05.

5. **Results and Discussion**

In this section we analyse student's improvement in conceptual understanding as well attitude both in terms of age and gender. We also compare our results of conceptual understanding with that of a 10-week long intervention and observe for any significant differences.

*5.1 An analysis of conceptual improvement and understanding*

Table 1: The average pre and post test scores of boys and girls from Year 7 to 10. The ratio between post and pre-test average indicates the improvement factor (IF) for each year level in terms of boys and girls

| No. of students (n) | Standard | Boys | | | | Girls | | | |
| --- | --- | --- | --- | --- | --- | --- | --- | --- | --- |
| | | No. of boys | Average pre-test score (%) | Average post-test score (%) | IF | No. of girls | Average pre-test score (%) | Average post-test score (%) | IF |
| 24 | Year 7 | 13 | 20 | 34 | 1.7 | 11 | 26 | 35 | 1.3 |
| 26 | Year 8 | 14 | 38 | 56 | 1.4 | 12 | 19 | 46 | 2.4 |
| 28 | Year 9 | 14 | 27 | 53 | 1.9 | 14 | 16 | 52 | 3.2 |
| 26 | Year 10 | 14 | 27 | 54 | 1.9 | 12 | 28 | 53 | 1.9 |
| N = 104 | | **Average Ratio = 1.7** | | | | **Average Ratio = 2.2** | | | |

In table 1 we analyze the performance of the students in terms of boys and girls for each year level. The average improvement factors of boys and girls were found to be 1.7 and 2.2 respectively.

In Figure 8(a)-8(d) we present histograms of students' performance before and after each program. We observe an improvement in students' conceptual understanding after each program. The average post-test conceptual scores of Year 8-10 students were greater than 50%, substantially higher than their pre-test scores. The average of Year 7 (who were talented language students) was ~35% substantially low in comparison to other three classes (who were talented science students). The number of students who scored more than 50% after each program are 2, 14, 17 and 15 for Year 7, 8, 9 and 10 respectively.

The figures indicate a significant difference between pre-test and post-test total conceptual scores. This implies that Einsteinian physics concepts are not covered well in school science curriculum and can be readily introduced across Year 7 to 10 through active learning methods.

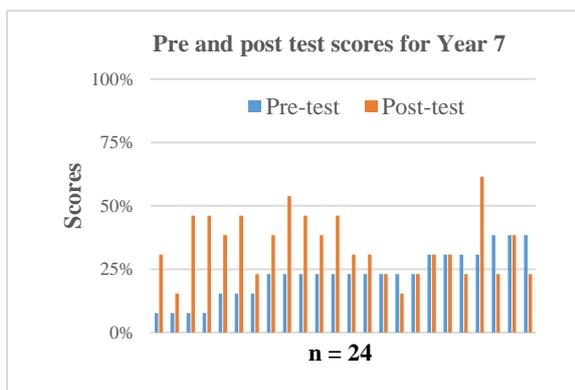
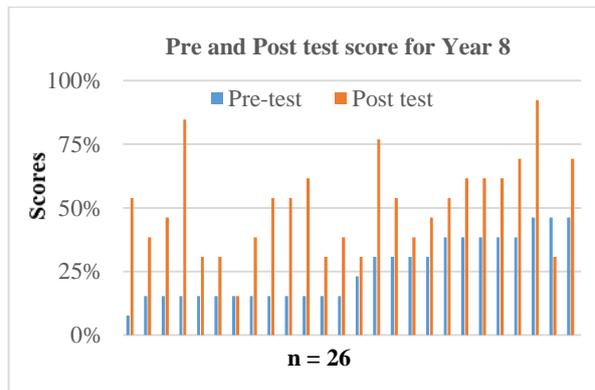

Figure 8(a): Pre and post test scores for 24 students of Year 7 arranged in ascending order of the pre-test results. We observe that five students do not undergo any change in his/her score and except four students who scored lower, all the students score significantly higher after the program. The average rise in score is by a factor of 1.5

Figure 8(b): Pre and post test scores for 26 students of Year 8 arranged in ascending order of pre-test results. We observe that except one student who scored lower and one whose scores were equal, all the students score significantly higher after the program. The average rise in score is by a factor of 1.9

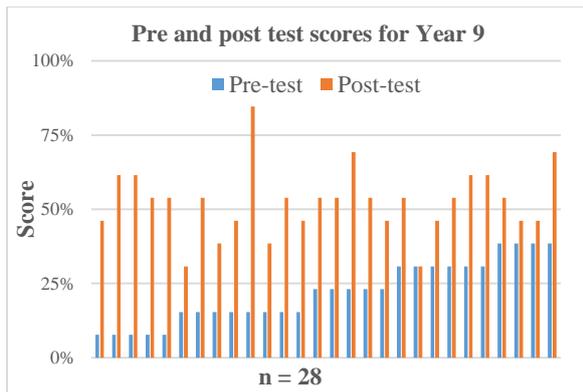 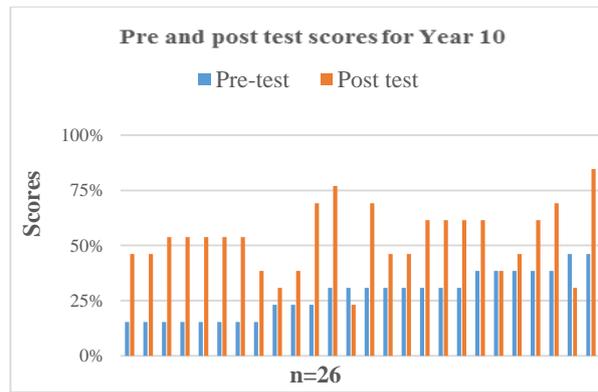

Figure 8(c): Pre and post test scores for 28 students of Year 9 arranged in ascending order of pre-test results. We observe that except one student who scored equal, all the students score significantly higher after the program. The average rise in score is by a factor of 2.6

Figure 8(d): Pre and post test scores for 26 students of Year 10 arranged in ascending order of pre-test results. We observe that except two students who scored less and one who scored equal, all the students score significantly higher after the program. The average rise in score is by a factor of 1.9

We found no correlation between pre and post-test score on overall understanding when analyzed in terms of age. A gender based analysis of the improvement is presented in Figure 9. The improvements of Year 8 and Year 9 girls were substantially higher than boys and Year 10 was marginally the same. Only in case of Year 7 the improvement factor was higher for boys. Finally, we analyze students' performance in terms of "core" and "derived" concepts presented in the Table 2.

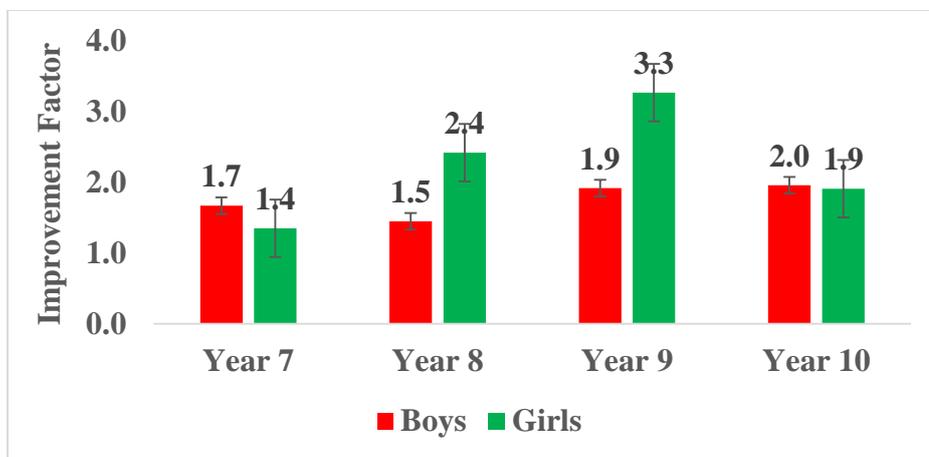

Figure 9: Gender based analysis on the improvement scores in conceptual understanding for each year level. The red and green histograms represent the improvement factors of boys and girls respectively. The improvements of Year 8 and Year 9 girls are substantially higher than boys and Year 10 is marginally the same. In Year 7 the improvement factor was higher for boys.

Table 2: An analysis of student's performance in terms of individual question as well as core and derived concepts

|  | Percentage of students answering correctly | | | | | | | |
|---|---|---|---|---|---|---|---|---|
| Questions | Year 7 | | Year 8 | | Year 9 | | Year 10 | |
|  | Pre (%) | Post (%) | Pre (%) | Post (%) | Pre (%) | Post (%) | Pre (%) | Post (%) |
| Q1. What is light? Explain in a few words. | 21 | 67 | 38 | 99 | 63 | 100 | 81 | 100 |
| Q2. Write down the names of three types of electromagnetic waves and explain what they are. | 12 | 25 | 50 | 85 | 29 | 75 | 81 | 93 |
| Q3. Can light exert forces on things? Explain your answer. | 25 | 83 | 19 | 99 | 10 | 99 | 21 | 99 |
| Q4. What aspect of light can cause uncertainty in measurement? | 0 | 19 | 0 | 37 | 0 | 12 | 7 | 45 |
| Q5. What aspect of light can be used to make very precise measurements? | 0 | 21 | 0 | 31 | 0 | 8 | 0 | 55 |
| Q6. Does light have energy? | 100 | 100 | 100 | 100 | 100 | 100 | 100 | 100 |
| Q7. Does energy have mass? | 33 | 55 | 27 | 33 | 27 | 79 | 21 | 43 |
| **Core concepts** | 23 | 75 | 28.5 | 99 | 36.5 | 99.5 | 51 | 99.5 |
| **Derived concepts** | 11 | 31.6 | 9 | 33.6 | 9 | 33.6 | 9.33 | 47.6 |
|  | $t < 0.05$ Statistically significant | | $t < 0.05$ Statistically significant | | $t < 0.05$ Statistically significant | | $t < 0.05$ Statistically significant | |

The answer to (Q6) were unanimously correct in every class in both pre and post-tests. We were surprised and impressed that not one student failed this question. Regarding $E=mc^2$, this concept was not covered explicitly in the program due to insufficient time. Thus the topic was not addressed well and we would expect no significant change. This raises a question: Why did scores for this question improve? It could be due to some students making the following logical connection: light has energy and light is made of photons. But photons have momentum, and momentum is associated with mass. Hence energy also has mass.

We conclude that the students were able to extrapolate the concept of $E=mc^2$ using their prior knowledge and their understanding of the derived concepts in the program.

After analyzing the core and the derived concepts, we observe that more than 95% Year 8-10 and ~75% of Year 7 students were able to answer the core concepts. We observe a near universal acceptance for the core concepts among students of Year 8-10. However, for the derived concepts there was comparatively weak acceptance with less than 35% of Year 7-9 and ~47% of Year 10 students being able to answer correctly.

As presented in the histograms below, there is a steady improvement as a function of age for the derived concepts. The highest improvement is observed for Year 10 students with an improvement factor of 5.1. The improvement for the core concept was the least for Year 10 students implying that in comparison to other classes, Year 10 students were more familiar with Einsteinian concepts prior to the program.

One significant conclusion is that the core ideas of Einsteinian physics can readily be introduced across Year 7-10. However, for more subtle concepts like the uncertainty principle, and how wavelength determines the ability to make sensitive measurements, the results are not strong. To understand this better, we will compare the results with a long intervention conducted previously with Year 9 students in Section 5.3.

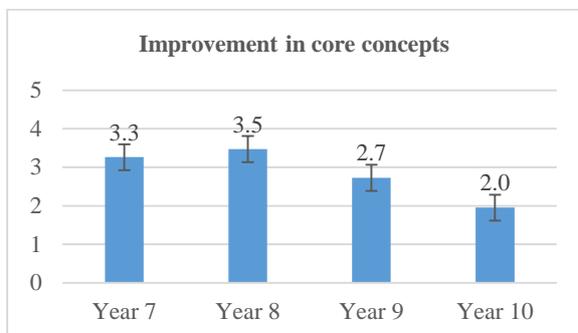

Figure 10(a): An analysis on the improvement scores in conceptual understanding of the core concepts based on each year level. The highest improvement factor is observed for Year 8. We observe no correlation in terms of age.

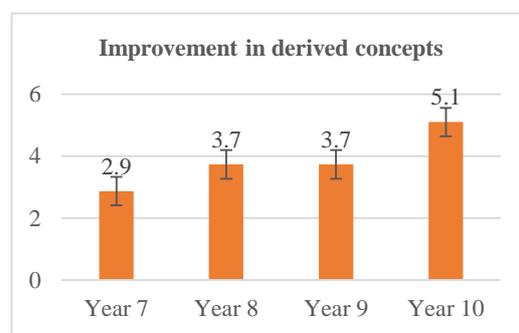

Figure 10(b): An analysis on the improvement scores in conceptual understanding of the derived concepts based on each year level. We observe a steady improvement in terms of age with the highest improvement factor of 5.1 for Year 10 students

## 5.2 An analysis of improvement in attitude

The following table summarizes the attitude test scores and improvement factors of boys and girls for each class year in terms of individual questions. The data shows that student attitudes to science (Q1) and to methods of learning (Q2) were generally very positively held before the trial, so that on average there was little room for improvement. However the inverted question on experiments (Q3) generally showed significant change. In the sections below we present an extensive analysis of the attitude of the students.

Table 3: An analysis of student's attitude in terms of gender and year level

|   | Questions | | Boys | | | | Girls | | | |
|---|---|---|---|---|---|---|---|---|---|---|
|   |   | Year | Pre Avg | Post Avg | IF | t-value p-value** | Pre Avg | Post Avg | IF | t-value p-value** |
| 1. | People need to understand science because it affects their lives. | Year 7 | 4.6 | 4.6 | 1.0 | t = 0 p =0.5 | 4.4 | 4.4 | 1.0 | t = 0 p = 0.5 |
|   |   | Year 8 | 4.6 | 4.6 | 1.0 | t =0.24 p = 0.4 | 3.9 | 4.1 | 1.1 | t =1.47 p = 0.08 |
|   |   | Year 9 | 3.6 | 4.0 | 1.1 | t = 0.8 p = 0.21 | 3.9 | 4.2 | 1.1 | t = 0.43 p = 0.33 |
|   |   | Year 10 | 4.4 | 4.8 | 1.1 | t = 2 p =0.04 | 4.4 | 4.4 | 1.0 | t = 1 p = 0.16 |
| 2. | I would prefer to find out why something happens by doing an experiment than by being told. | Year 7 | 4.7 | 4.7 | 1.0 | t = 0.8 p = 0.21 | 3.8 | 4.7 | 1.2 | t = 1.77 p = 0.05 |
|   |   | Year 8 | 3.7 | 4.2 | 1.1 | t = 0.66 p = 0.2 | 4.3 | 4.3 | 1.0 | t = 0.82 p = 0.21 |
|   |   | Year 9 | 3.3 | 4.0 | 1.2 | t = 0.32 p = 0.37 | 3.1 | 3.7 | 1.2 | t = 0.71 p = 0.24 |

| | | | | | | | | | |
|---|---|---|---|---|---|---|---|---|---|
| | | Year 10 | 4.5 | 4.5 | 1.0 | t = 1.51<br>p = 0.08 | 2.3 | 3.5 | 1.5 | t = 1.72<br>p = 0.05 |
| 3. | I would rather agree with other people than do an experiment to find out for myself.* | Year 7 | 2.6 | 4 | 1.5 | t = 0.29<br>p = 0.38 | 3.1 | 3.8 | 1.2 | t = 1.47<br>p = 0.06 |
| | | Year 8 | 2.2 | 3.9 | 1.8 | t = 1.62<br>p = 0.06 | 4.0 | 4.0 | 1.0 | t = 1.47<br>p = 0.08 |
| | | Year 9 | 3.5 | 3.5 | 1.0 | t = 0.8<br>p = 0.21 | 3.5 | 3.5 | 1.0 | t = 0.45<br>p = 0.32 |
| | | Year 10 | 3.3 | 4.6 | 1.4 | t = 0.8<br>p = 0.22 | 2.5 | 4.1 | 1.6 | t = 3.25<br>p= 0.005 |
| 4. | I would like to have a career in science. | Year 7 | 3.5 | 3.6 | 1.0 | t = 0.71<br>p = 0.24 | 3.6 | 3.6 | 1.0 | t = 0.71<br>p = 0.24 |
| | | Year 8 | 3.1 | 3.7 | 1.2 | t = 0.89<br>p = 0.19 | 3.6 | 3.6 | 1.0 | t = 1.47<br>p = 0.08 |
| | | Year 9 | 2.9 | 3.4 | 1.2 | t = 0.8<br>p = 0.21 | 3.8 | 3.8 | 1.0 | t = 1.38<br>p = 0.09 |
| | | Year 10 | 4.5 | 4.5 | 1.0 | t = 1.15<br>p = 0.14 | 4.6 | 4.6 | 1.0 | t = 1.16<br>p = 0.12 |
| 5. | Science classes teach me new things which are interesting | Year 7 | 3.9 | 4.3 | 1.1 | t = 0.56<br>p = 0.29 | 4.2 | 4.1 | 1.0 | t = 1.39<br>p = 0.09 |
| | | Year 8 | 3 | 4.5 | 1.5 | t = 1.3<br>p = 0.11 | 4.4 | 4.0 | 0.9 | t = 2.73<br>p= 0.003 |
| | | Year 9 | 4 | 4.4 | 1.1 | t = 1.39<br>p = 0.09 | 3.9 | 3.9 | 1.0 | t = 0.36<br>p = 0.36 |
| | | Year 10 | 4.2 | 4.2 | 1.0 | t = 1<br>p = 0.17 | 3.8 | 4.3 | 1.1 | t = 1.16<br>p = 0.13 |
| 6. | Science classes would be more interesting if we learnt modern topics. | Year 7 | 3.8 | 3.8 | 1.0 | t = 0<br>p = 0.5 | 3.0 | 3.5 | 1.2 | t = 0<br>p = 0.5 |
| | | Year 8 | 3.1 | 3.8 | 1.2 | t = 0.55<br>p = 0.29 | 3.7 | 3.7 | 1.0 | t = 0.48<br>p = 0.31 |
| | | Year 9 | 2.7 | 3.6 | 1.3 | t = 1<br>p = 0.16 | 2.7 | 3.8 | 1.4 | t = 0.29<br>p = 0.38 |
| | | Year 10 | 3.4 | 3.4 | 1.0 | t = 0.8<br>p = 0.22 | 2.6 | 4.0 | 1.5 | t = 1.44<br>p = 0.08 |
| 7. | Science is only for smarter people.* | Year 7 | 3 | 3.9 | 1.3 | t = 0.63<br>p = 0.26 | 4.1 | 4.2 | 1.0 | t = 0.24<br>p = 0.4 |
| | | Year 8 | 3.8 | 3.8 | 1.0 | t = 1.5<br>p = 0.08 | 3.9 | 4.2 | 1.1 | t = 1.14<br>p = 0.13 |
| | | Year 9 | 3.1 | 4.2 | 1.3 | t = 2.24<br>p = 0.02 | 2.6 | 4.4 | 1.7 | t = 1<br>p = 0.16 |
| | | Year 10 | 2.4 | 3.2 | 1.3 | t = 1<br>p = 0.17 | 3.6 | 3.9 | 1.1 | t = 1.4<br>p = 0.08 |

*Score are inverted for analysis.
**t = 0.05

1. *People need to understand science because it affects their lives.*

The very high scores show that this question is not a useful measure of attitude for academically talented students. The clear ceiling-effect for this question demonstrates that virtually all students came to the program with a very positive attitude to the importance of science. Slight improvement for Year 9 and Year 10 boys are not significant.

2. *I would prefer to find out why something happens by doing an experiment than by being told.*
3. *I would rather agree with other people than do an experiment to find out for myself.*

Questions 2 and 3 test student interest in an experimental approach to learning. The very high scores in 2 demonstrate that students entered the trial with a very positive attitude to experiments and there was little room for improvement. In spite of this a significant increase in score was observed across all year levels. However the inverted Q3 testing the same idea showed substantial changes. There were varied responses in support of learning by experiment, including substantial strong responses for Year 8 boys and Year 10 girls. On average there is a preference for activity-based learning by all the classes tested.

*4. I would like to have a career in science.*

This question indicates the program had minimal effect on career choices, contrary to the results reported by S. Laursen et al. [15] There was small effect for Year 8 and Year 9 boys towards career choices. The high ceiling effect resulted in comparatively less improvement in Year 10 students and no improvement for Year 7 presumably because they were selected language students.

*5. Science classes teach me new things which are interesting.*

This question tests students' enjoyment of science. The high pre-scores again show that students are already have an extremely positive attitude to science, which is probably an outcome of the widely recognized high quality science teaching at the particular school. However we do observe an improvement in attitude for Year 7-9 boys and Year 10 girls. The highest improvement was for Year 8 boys whose improvement factor was 1.5. This suggests that the program did spark some extra interest in this group.

*6. Science classes would be more interesting if we learnt modern topics.*

Significant improvement factors in Year 9 and Year 10 indicate that students found the new topics interesting. Their relatively low pre-scores indicate that they did not enter the program with negative views that their current curriculum is weak in modern topics. Students agree that modern topics are interesting and are not included extensively in their curriculum. The highest improvement is for Year 10 girls whose improvement factor was 1.5. This perhaps indicates that students became aware that modern topics are not covered extensively in their syllabus through the program itself.

*7. Science is only for smarter people.*

This question tests stereotypical views that science is a subject for nerds and smarter people. A moderate improvement in attitude was observed in three classes. The highest improvement was for Year 9 girls whose improvement factor was 1.7. The response to this question suggests that the program was able to demystify science among the students.

Overall the attitude test shows that the combination of activity-based learning and meeting real researchers created a positive attitudinal change in students. While there was scatter between classes, there was no particular age dependence for attitude.

In this program, 77% of the total students were gifted and talented in science who scored very high in the pre-tests which resulted in the clear ceiling-effects observed. The questions used in the program could be more useful in testing attitudes of less motivated and lower achieving students. This attitude questionnaire also highlights the difficulty in designing questionnaires suitable for use across a broad range of students.

It is interesting that despite lower knowledge scores, we observed an improved attitude in students of Year 7 which were quite consistent with the improvement of Year 8-10. This indicates that Einsteinian physics interventions could induce positive changes in students' attitude, regardless of their academic level or age group which is an interesting case to investigate further.

When analyzed overall in terms of gender, the improvement in attitude was moderate compared to conceptual improvement.

*5.3. A comparison with 10-week long intervention program*

A 10-week program on Einsteinian physics was conducted by Kaur et al. in 2013 and 2014 with academically talented Year 9 students. [14] The 20-lesson program focused on teaching general relativity and quantum physics using similar models and analogies to those used here. With substantially more time available there was enough time for ideas to be assimilated and students could consolidate the concepts with worksheets that followed up each session. Students' response were analyzed using similar pre and post-test questionnaires as used in the program described above.

The conceptual questionnaires were designed to measure the intake of the core understanding of Einsteinian physics by asking questions such as: "Can two parallel lines meet?" "What do you understand by the term light?" The attitudinal questions were designed to test students' attitude by asking questions based on students' interest in physics, relevance of Einsteinian physics etc.

Some of the activities used in the one day program were similar to the 20-lesson program. For example in both the programs, nerf-gun projectiles were used as an analogy to describe the photon momentum effect. Also similar conceptual questions (for example "What is light") were used in both the programs to measure the students' core understanding of Einsteinian physics.

Therefore, it is quite reasonable to compare the results of the two programs and observe for differences. The results on conceptual understanding from the two programs have been presented in the table below:

Table 4: An analysis of student's understanding in terms of duration of intervention

|  | One day intervention with Year 7-10 | | | | | | | | 10-week intervention with Year 9 | | |
|---|---|---|---|---|---|---|---|---|---|---|---|
|  | Percentage of students who gave correct answer | | | | | | | | Percentage of students who gave correct answer | | |
|  | Year 7 | | Year 8 | | Year 9 | | Year 10 | | Year 9 | | |
|  | Pre (%) | Post (%) | Pre (%) | Post (%) | Pre (%) | Post (%) | Pre (%) | Post (%) | Year of program | Pre (%) | Post (%) |
| **Core concepts** | 23 | 75 | 28.5 | 99 | 36.5 | 99.5 | 51 | 99.5 | **2013** | 27 | 73 |
| **Derived concepts** | 11 | 31.6 | 9 | 33.6 | 9 | 33.6 | 9.33 | 47.6 | **2014** | 23 | 91 |

We observe that in the 10-week program, students' average post-test score is 73% and 91% in Year 2013 and 2014 respectively. According to Kaur et al. this difference occurred due to the following two factors: (a) the presenter's skill significantly improved in 2014 through the experience of the first year, (b) In 2014 students undertook a mid-program test which may have helped to reinforce their learning.

In the one day program the average post-test score for core concepts was more than 95% for Year 8-10 students and 75% for Year 7 students. This is comparable to the 10-week program. However, the students score significantly less when we consider the derived concepts. The students of Year 7-9 scored within 31-34% and the students of Year 10 scored 47% in the derived concepts.

One interesting observation from the pre and post test results was that the scores were independent of each other in both cases except Year 7 in the one day program. This suggest that prior knowledge is not a factor for comprehending Einsteinian physics concepts. Moreover students were able to understand the concepts irrespective of their age or duration of the program. Above results also suggest that short programs are not suitable to introduce more subtle concepts of Einsteinian physics such as photon momentum can cause uncertainty in measurement or wavelength of light sets the scale of measurement. Therefore, we can conclude that long interventions can be used to focus on the derived concepts while short intervention will be suitable for core concepts.

## 6. Conclusion

From Figure 10 (a) & (b) we can see a clear age dependence for the derived concepts while core concepts are independent of any age group. Students understand the derived concepts in long intervention which suggest that the scope for short interventions need to be reduced and the extended program can lead to a broad literacy of the Einsteinian concepts.

It was Year 10 whose conceptual understanding of the derived concepts was the highest. This suggests that it could be better to focus on core concepts at earlier ages, and introduce the derived concepts later. This could be tested by covering similar material in greater depth through longer interventions, to determine whether it is age or exposure time that determines understanding of more subtle and derived concepts.

A lower response from the non-science students suggest that program need to be adapted to support weaker students. The flat response of post-test results as a function of pre-test scores as reported by Kaur et al. [14] suggest that talent (as measured by pre-test score) is not the determining issue for introducing Einsteinian physics. This apparent contradiction need to be resolved with further studies. We are trying to resolve this issue by conducting a 10-lesson program with different talent stream of Year 7 classes.

There was an attitudinal difference observed in both the program. When averaged over all the students the conceptual mean scores improve by a factor of 1.7 and 2.2 for boys and girls respectively. In spite of strong ceiling effect, the attitudinal mean scores improve by a factor of 1.17 and 1.14 for boys and girls respectively. It was interesting to see that in two out of four cases, girl's improvement in understanding substantially exceed the boys' (in one case it was almost equal), although the girls started with comparatively lower score before the intervention.

All the above results support previous findings which indicate that Einsteinian physics is readily accepted by high school students. There is significant evidence that girls respond more positively compared to boys. We cannot determine whether this is due to the nature of the material or the activity based learning, which has previously been shown to encourage female improvement. [33] Hence this program can be used as an effective case study to create longer interventions for further implementation of Einsteinian physics.

**Acknowledgement**

This research was supported by a grant from the Australian Research Council, the Gravity Discovery Centre and the Graham Polly Farmer Foundation. The authors are grateful to Yohanes Sudarmo Dua, Si Yu, John Moschilla, Andrew Burrell, Vladimir Bossilkov, Carl Blair, Joshua McCann, Ju Li, all the relief teachers and students who participated in this study.

**Appendix 1**

## ROLE PLAY SCRIPT
### The story of light
Electromagnetic Waves, Light, Photons and Gravitational Waves

**(Participants: Narrator, Heinrich Hertz, Journalist, Albert Einstein & Richard Feynman)**

*Narrator*: "*We are going back to a time when your great-great grandparents were young! The year is 1886, and we are in Heinrich Hertz's laboratory in modern-day Germany*"

Heinrich Hertz (standing over his laboratory apparatus) [triumphantly]: "*Wow! I have discovered electromagnetic waves. I've just sent an invisible wave across this table!*"

Journalist (with a pen and notepad in his hand) [sceptically]: "*That's all well and good Mr. Hertz, but what is the use of your discovery?*"

Heinrich Hertz [proudly, loudly]: "*It is of no use whatsoever. I was just trying to prove what Maxwell asserted and he is right.*"

Journalist [taken aback]: "Oh! ….. Ok….. *So what are these electromagnetic waves?*"

Heinrich Hertz: "*I have no idea! All I know is that we have these mysterious electromagnetic waves that we cannot see with the naked eye………... But they are there.*"

Narrator: Who do you agree with? Was Hertz correct in saying his discovery was of no use whatsoever? What do you think?

**(Class discussion)**

Narrator: "*We are back in Hertz's laboratory, one year later in 1887*"

Journalist (holding the notepad) [still sceptical]: "*So Mr Hertz, I hear that you have discovered something weird?*"

Heinrich Hertz [proudly]: "*Yes, I have!... I've discovered something really mysterious.*"

Journalist: "*And what's that Mr. Hertz?*"

Heinrich Hertz: "*I found that shining light on metals causes electric currents to flow through empty space.*"

Journalist: "*Wow…that is weird.*"

Heinrich Hertz: "*Yes it is. None of my friends can explain it.*"

**(Interlude)**

Narrator: "*Now it is 18 years later. We have moved to a little patent-office in Switzerland. The year is 1905.*"

Albert Einstein: "Eureka! *I have solved Hertz's mystery!*

Journalist: "*What is the solution Mr. Einstein?*"

Albert Einstein [triumphantly]: "*Light comes like little bullets*"

Journalist [perplexed]: "*But I'm confused Mr. Einstein. Hertz told me that it's a wave*"

Albert Einstein: "*Well, everyone used to believe that light was a wave. That's what they taught me in school*"

Journalist: "*I'm going to need a second opinion.*"

Narrator: *"The journalist is looking for someone to ask. His search takes more than half a century. It's the nineteen sixties, and he's a bit old now."*

Journalist [old and tired]: *"Hey..Mr. Feynman. They tell me that you're the greatest physicist alive today! Einstein told me that light comes in little bullets called photons. Tell me…. do you agree with Einstein?"*

Feynman: *Yes. Definitely, I agree with him"*

*"I wrote about it in one of my most famous book". This is what I wrote:* [21]

*"I want to emphasise that light comes in this form – particles. It is very important to know that light behaves like particles, especially for those of you who have gone to school, where you were probably told something about light behaving like waves"*

Einstein [from beyond the grave. i.e. off-stage]: *"And what about my very best discovery, Mr Feynman. I predicted that there was another sort of wave: gravitational waves."*

Feynman: *"Sorry not to mention that Albert. Indeed it was a great discovery". Did you know it was me who first proved that those waves could be detected?"*

Narrator*: "And on the 14th of September 2015 those waves were detected for the very first time".*

***end of role play***

**Appendix 2**

**MEASURING THICKNESS OF HUMAN HAIR USING DIFFRACTION**

The diameter of a hair can be calculated by measuring the spacing between successive bands in the interference pattern and employing the following simple formula:

$$d = \frac{\lambda L}{x}$$

where d is the hair diameter, $\lambda$ is the wavelength of the laser light, L is the distance between hair and screen, and x is the distance between successive light/dark bands. We simplify the formula for high school students by giving them a numerical value. The distance between the hair and the screen is 2.5 m and the wavelength of green light is 532 nm, as a result the formula simplifies to the following numerical value after doing the calculation:

$$d \text{ (in micron)} = \frac{113}{Fringe\ spacing\ (in\ cm)}$$

**References**

[1] http://sitn.hms.harvard.edu/flash/2016/ligo-and-gravitational-waves-discovery-of-the-century/ Retrieved on February 13, 2018

[2] https://www.ligo.caltech.edu/page/what-are-gw Retrieved on February 14, 2018

[3] Merger B. P. Abbott et al. (2016). "Observation of Gravitational Waves from a Binary Black Hole". *Phys. Rev. Lett.* **116**

[4] https://phys.org/news/2017-11-physicists-rapid-bounding-gravity.html Retrieved on 18th February, 2018 Retrieved on February 18, 2018

[5] Two kinds of waves from a neutron-star smashup Physics Today 70, 12, 19 (2017); https://doi.org/10.1063/PT.3.3783 Retrieved on February 13, 2018

[6] Bungum, B., et al. (2015). *"ReleQuant – Improving teaching and learning in quantum physics through educational design research"*. *Nordic Studies in Science Education* Vol **11**, No. 2

[7] https://education.gov.scot/scottish-education-system/policy-for-scottish-education/policy-drivers/cfe-(building-from-the-statement-appendix-incl-btc1-5)/What%20is%20Curriculum%20for%20Excellence? Retrieved on May 20, 2018

[8] https://keynote.conferenceservices.net/resources/444/5233/pdf/ESERA2017_0583_paper pdf Retrieved on May 20, 2018

[9] https://www.nsf.gov/bfa/lfo/seminars/pub/ligo_educationoutreach.pdf Retrieved on February 22, 2018

[10] Baldy, E. (2007). "A New Educational Perspective for Teaching Gravity". *International Journal of Science Education*, **29**, p.1767-1788.

[11] Kaur, T. et al. (2017). "Teaching Einsteinian physics at schools: part 3, review of research outcomes". *European Journal of Physics Education* **52** (6).

[12] Kaur, T. et al. (2017). "Teaching Einsteinian physics at schools: part 1, models and analogies for Relativity". *European Journal of Physics Education* **52** (6).

[13] Kaur, T. et al. (2017). "Teaching Einsteinian physics at schools: part 2, models and analogies for quantum physics". *European Journal of Physics Education* **52** (6).

[14] Kaur, T. et al. (2018). "Evaluation of 14 to 15 Year Old Students' Understanding and Attitude towards Learning Einsteinian Physics". Retrieved from https://arxiv.org/abs/1712.02063 on May 25, 2018

[15] Laursen, S, et al. (2007). "What Good Is a Scientist in the Classroom? Participant Outcomes and Program Design Features for a Short-Duration Science Outreach Intervention in K–12 Classrooms". *CBE—Life Sciences Education,* **Vol. 6**, 49–64, Spring-2007

[16] Wynarczyk, P. & Hale, S. (2009). "Improving take up of science and technology subjects in schools and colleges: A synthesis review", Report prepared for the Economic and Social Research Council (ESRC) and the Department for Children, Schools and Families

[17] Hassan, G (2008). "Attitudes toward science among Australian tertiary and secondary school students". *Research in Science & Technological Education*, **26:2**, 129-147, DOI: 10.1080/02635140802034762

[18] Jones, L. R., Mullis, I. V. S., Raizen, S. A., Weiss, I. R. and Weston, E. A. (1992). "The 1990 science report card". *Washington, DC: Educational Testing Service.*

[19] https://www.ligo.caltech.edu/video/ligo20160211v1

[20] Aspden, R. S. & Padgett, M. J. (2016). "Video recording true single-photon double-slit Interference" *American Journal of Physics* **84**, 671.

[21] Feynman, R. (1985). QED: The Strange Theory of Light and Matter (Alix G. Mautner Memorial Lectures) (New Jersey: Princeton University Press. p.15.

[22] Taylor, G. (1909). "Interference fringes with feeble light". *Proceedings of the Cambridge Philosophical Society* **15**, p. 114-15.

[23] Tinsley, J. N. et al. (2016), *Nature Communication*. **7**, 12172.

[24] https://www.youtube.com/watch?v=I9Ab8BLW3kA Retrieved on September 22, 2017

[25] Prince, M. (2004). "Does active learning work? A review of the research". *J. Eng. Educ.* **93**, 223.

[26] Hertz and Lenard's Observations of The Photoelectric Effect. Retrieved from http://byjus.com/physics/hertz-lenard-observations/ on September 20, 2017.

[27] Arons, A.B. & Peppard, M. B. (1965). "Einstein's Proposal of the Photon Concept–a Translation of the Annalen der Physik Paper of 1905". *American Journal of Physics* **33**, 367

[28] Interference of a single photon Retrieved from http://phy-page imac.princeton.edu/~page/single_photon.html on September 22, 2017 https://www.youtube.com/watch?v=MbLzh1Y9POQ

[29] https://en.wikipedia.org/wiki/Phasor#/media/File:Unfasor.gif Retrieved on May 23, 2017

[30] Barish B.C. & Weiss, R. (1999). "LIGO and the detection of gravitational waves". *Phys. Today* **52**, 10, 44–50.

[31] https://www.educ.cam.ac.uk/research/projects/episteme/epiSTEMeScienceAttitude Questionnaire.pdf Retrieved on April 21, 2017

[32] https://www.iop.org/education/teacher/support/girls_physics/action/file_41604.doc Retrieved on April 21, 2017

[33] Kaur, T. et al. (2018). "Gender response to Einsteinian physics interventions in School". Retrieved from https://arxiv.org/abs/1712.06323 on May 25, 2018